\documentclass{grstyle}
\usepackage{graphicx}

\begin{document}

\begin{frontmatter}

\title{Preliminary Study of the Feasibility of a Non Crystalline Positron Emission Tomography Using a Suspension of
Superheated Superconducting Grains (SSG) 
in High Density Dielectric Matrix (HDDM) as Detector}

\author[CNRSGPS,IN2P3]{Roger Bru\`ere Dawson},
\author[IN2P3,CNRSIDRIS]{Jacques Maillard }\ead{maillard@idris.fr},
\author[P6SA]{G\'erard Maurel},
\author[P6]{Jorge Silva},
\author[CNRSGPS]{Georges Waysand}

\address[CNRSGPS]{GPS UMR 7588, Universities Paris VI and Paris VII, 
4 place Jussieu, 75005 Paris France}
\address[IN2P3]{IN2P3, 3 rue Michel Ange, 75794 Paris Cedex 16}
\address[CNRSIDRIS]{CNRS IDRIS, Bt 506, BP167 91403 Orsay Cedex France}
\address[P6SA]{CHU Hopital Saint Antoine, Paris VI, Facult\'e de M\'edecine, 27 rue de Chaligny 75012 Paris France}
\address[P6]{Universit\'e Paris VI, 4 place Jussieu, 75005, Paris, France}


\begin{abstract} 
Suspensions of superheated superconducting grains are a detecting composite material. Each grain in the supension  is a microcalorimeter with an energy threshold defined by its equatorial magnetic field for a given temperature. The higher the matrix density, the larger the gamma stopping power.
For several years, cylindrical cells of such suspensions about 2 cm long and 0.4 mm in diameter can be read out in real time. As a result, using two independent cells, one can record a time coincidence between them. This could be potentially very useful for positron cameras where two diametrically opposite cells are simultaneously knocked by 511 keV gammas. This paper, based on the state of art in SSG in high density matrix, discusses such a feasibility.
\end{abstract}

\end{frontmatter}
\section{Positron Emission Tomography (PET) and Suspensions of Superheated 
Superconducting Grains in High Density Matrix}  

Superheated superconducting grains detectors (SSG) are made from a suspension 
of microspheres of type I superconductor. Kept at a temperature lower than 
the superconducting critical one and in a DC magnetic field larger than the 
thermodynamical critical value at that temperature, each grain remains in a 
superconducting state, which is indeed a superheated superconducting state 
(this is not valid for a bulk superconductor because of surface defects acting as permanent nucleation centers). This superheated state can be
broken down, either by ramping up the magnetic field, or by heating the 
microsphere (by energy deposition). In both cases, the perfect diamagnetism 
associated with the superconducting state disappears suddenly.  The resulting 
local magnetic flux variation can be detected in real time by a pick-up coil 
surrounding the whole suspension.
 Although good type I superconductors are : Al, Zn, Ga, Ge, Cd, In, Sn, Hg,Tl, the 
most commonly used metal is tin because of its characteristic values for the 
superconducting state: Tc = 3.72K just below the 4, 2K for liquid helium under 
atmospheric pressure, Hc(0)  (at zero K)  =302 Gauss Hsh(0) =600 Gauss. Low temperature techniques are no more a problem in medical environment: MRI magnets are immersed in tens of litres of liquid helium and on the other hand pulse tube system can provide liquid helium in closed loop systems. Non 
toxicity and friendly metallurgy of microspheres are secondary reasons for 
that choice. Usually the microspheres are randomly but homogeneously dispersed 
in paraffin wax (reference \cite{a_waysan1}). As a result such a suspension has a low stopping power 
for 511 keV gamma since the filling factor in tin grains cannot be larger 
than 0.1 to avoid percolation between grains. For positron camera, it is 
possible to replace wax by litharge (PbO), the composite of tin grains and 
PbO is therefore a suspension of SSG in High Density Dielectric Matrix (HDDM). 
Because of the presence of high Z element in such a composite one has to take 
into account not only the ionization process but also the photoelectric 
effect under 511 keV irradiation.
In recent years the sensitivity of the sudden flux change read-out has been enhanced with the use of a 
preamplifier  with an HEMFET input stage (reference \cite{b_BruereDawson}) cooled in liquid helium. 
We assume throughout this paper that the sensitivity of the preamplifier is high enough to 
read a single  grain larger than 10$\mu$m in diameter in the pixel defined below or in an equivalent volume.
For the following : 
\begin{itemize}
\item[1-] We recall the basics physics of SSG.
\item[2-] We describe the detecting system with SSG-HDDM, with its main parameters,
technical limitations and constraints limiting ourselves to a single cell of the previous system.
\item[3-] Using GEANT (reference \cite{c_Geant}) simulation program we study its behavior under 511keV irradiation. 
These results allow us to estimate the read-out performance under positron irradiation.
\item[4-] Finally we discuss the requirements for a completion of such a system.
\end{itemize}

\section{Basic Physics and Geometry of the elementary pixel of a Positron Camera with HDDM-SSG }

\subsection{ Basic physics}

\begin{figure}[!htb]
 \begin{center}
 \resizebox{!}{8cm}{
 \includegraphics{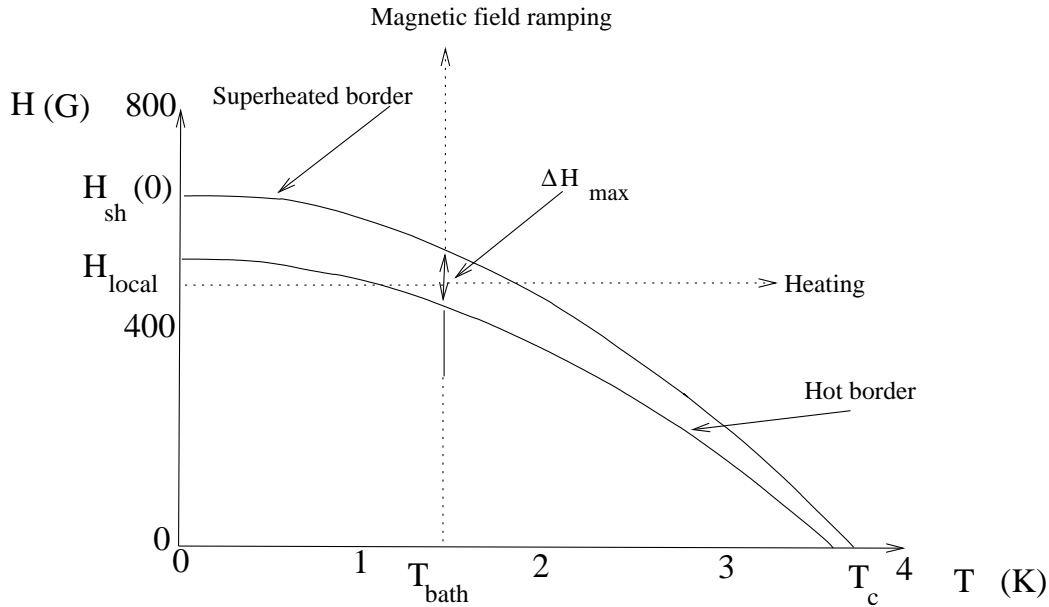}}
 \caption{Phase  diagram  \label{diagphase6}}
 \end{center}
\end{figure}


In a phase diagram (T, H) (reference \cite{d_helene} page 2, see figure\ref{diagphase6}) one grain of the suspension can be
represented by one point (H$_{local}$, T$_{bath}$). T$_{bath}$ is the
bath temperature in which the suspension is immersed.  H$_{local}$ is
the value of the magnetic field at the equator of the grain or its
surface maximum if the grain is not perfectly spherical. H$_{local}$
is different from H$_{applied}$ because of diamagnetic interactions between
grains. Diamagnetism is the magnetic field expulsion from the volume
of the superconducting material (here the grain), as a result from grain to grain
there are field variations induced by these magnetic flux expulsions. 
Therefore in the phase diagram a suspension of grains at a given
temperature T$_{bath}$ is represented by a set of points along the vertical
line T=T$_{bath}$.  If these grains are numerous, the suspension can
be represented by a vertical segment : each point of this segment
represents one grain and only one. There are two main paths for a
grain to undergo a change of state : either by heating or by magnetic
field ramping up. Heating can be produced by energy deposition, if enough
energy is deposited one has a thermal nucleation mechanism.  In the
case of magnetic field ramping up one speaks of magnetic nucleation. It
is the distance from the representative point in the phase diagram to
the superheated border which defines the amount of energy necessary
for thermal nucleation. It has been demonstrated that the heating of a
superheated grain under energy deposition is, after a very short time,
($10^{-10}$-$10^{-12}$ second), homogeneous.
     
   In principle, since the diamagnetic interactions between
microspheres are long range interactions, each transition of a
microsphere to the normal (resistive state) modifies the magnetic
field sensed by each grain of the suspension. However, if only a small
fraction of the grains undergo a transition one can safely consider
in first approximation that these modifications are negligible. It is only when a large
fraction of the grains have changed state that one must take into
account the modification of diamagnetic interactions. This problem has
been analytically solved by Geigenm\"{u}ller and Mazur (reference \cite{e_Geigenmuller1}): given the
geometrical positions of the remaining superconducting grains, the
magnetic field at the equator of each of them is computable (reference \cite{f_Geigenmuller2}).
More recently it has been discussed by Pe\~{n}aranda et al (reference \cite{g_Penaranda}).

\subsection{ Geometry of a single pixel}

As a starting point to evaluate which kind of benefit one could expect
from such a system as a positron camera, we take the most conservative parameters for the
elementary cell : 
\begin{description}
\item
Geometry: a cylinder 0.78 cm long ; 0.4 cm in diameter
\item
Temperature: about 200mK 
\item
Magnetic Field: Hc (0,2K)= 300 Gauss, Hsh (0,2K)= 600 Gauss. 
\end{description}
It is assumed that each cell is a suspension with a filling factor of 0.1 for
tin microspheres of respectively 7, 8 or 10$\mu$m  diameter.  For larger filling factor
contacts or quasi contacts between grains create strong diamagnetic
interactions resulting in local fields stronger than the theoretical
superheated critical magnetic field~: the grains are no longer
superconducting. 
The energy loss of the particle being proportional to the radius of the microsphere of the trajectory, and the heating of the microsphere proportional to the inverse of the cubic of radius, the heating decreases inversely to the square of the radius. The energy thresholds  decrease with the radius of the microsphere.    

Due to the modularity of the system, we can take one of these cells to simulate
the behaviour under 511 keV irradiation of the full detector. Our
tool for simulation is GEANT 321 (reference \cite{c_Geant}. This program takes into account all the possible interactions of an impinging photon on the cell: ionisation and
radiation including secondary effects. It is a step by step
simulation: the trajectory of the particle is incremented step by step
in a minute quantity depending on one hand on the particle nature,
lifetime and momentum, and on the other hand on the crossed media
(chemical composition, density and surrounding boundaries).  To
simulate the whole pixel needs too much memory space, therefore we
consider a fraction of it  along the
irradiation axis and a diameter of 0.1 cm. GEANT is unable to deal at
once with such a large number of grains; therefore, along the axis we
divide the cylinder into sub cylinders. The final simulation represents
a volume of 8 mm3. In GEANT one is free to choose the minimum step: it
has been taken here as a fraction of a micrometer. For each step GEANT
gives the energy loss in that step, the eventual gamma interaction,
the deposited energy per grain and finally the number of crossed
granules with energy deposition.

\section{Deposited energies in the suspension}

\begin{figure}[!htb]
 \begin{center}
 \resizebox{!}{3cm}{
 \includegraphics{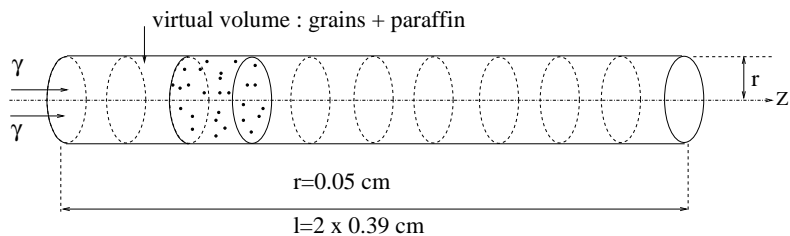}}
 \caption{Geometry of simulation  \label{figure_2}}
 \end{center}
\end{figure}

The gamma beam is parallel to the cylinder axis, arbitrarily chosen as
the Z axis (\ref{figure_2}).  $10^5$ photons are sent onto the detector
which is a suspension of monosized tin microspheres either 7, 8, or 10
$\mu$m in diameter, representing 10\% of the volume cylinder. The
microspheres are immersed in a high density matrix made of litharge
(Pb0, density =9.35). Such a detector irradiated by 511keV gammas has
a stopping power of about 37\% : Simulation gives that out of 100 000 photons impinging on
the cylinder, 36713 interact by electron production
either by Compton or by photoelectric effect. As we will see the
energy deposition in both processes occurs in the vicinity of the spot
of materialisation of the impinging photon.

\subsection{Distribution of the interaction points:}
\begin{figure}[!htb]
 \begin{center}
 \resizebox{!}{7cm}{
 \includegraphics{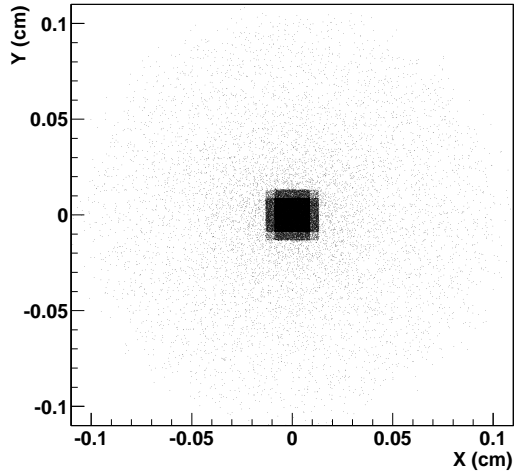}}
 \caption{xy projection of $\gamma$ impact: its dispersion is one order of 
 magnitude smaller than the radius of the simulated cylinder  \label{figure_3}}
 \end{center}
\end{figure}

Figure \ref{figure_3} represents the projections (X and Y) on a diametrical plane of the interaction spot of each 
interacting gamma in the whole cylinder. The aperture of the gamma beam is reduced to 
10$\mu$m around the Z axis. In this representation one does not care if the interaction 
takes place in a microsphere or in the matrix. 
One can see that the diameter of the virtual volume is large enough to include all the diffused
events.

\subsection{Distribution of crossed microspheres :}
\begin{figure}[!htb]
 \begin{center}
 \resizebox{!}{7cm}{
 \includegraphics{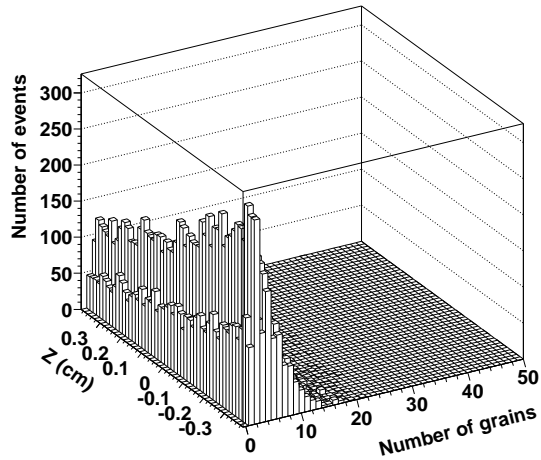}}
 \caption{Number of crossed grains versus the depth Z: the number of grains does not depend of Z \label{figure_4}}
 \end{center}
\end{figure}

As long as the photon does not interact with the pixel there is no energy
deposition. This deposition takes place only along the secondary trajectories.
The energy loss useful for read-out of a photon impact occurs only in the grains.  We reserve the 
denomination "crossed grains" or " crossed microspheres" to these specific microspheres 
in which there is an energy deposition. Figure \ref{figure_4} represents the number of crossed grains 
per incoming photon as a function of the depth Z  in the cylinder. As expected, since 
the mean free path of electrons is shorter than the mean free path for photons one can 
see that the number of crossed granules is depth independent of the photon impact. 
As a result, the energy loss after a photon interaction is independent of the point where it has occurred.

\subsection{Distribution of crossed grains with energy of gamma: }
\begin{figure}[!htb]
 \begin{center}
 \resizebox{!}{7cm}{
 \includegraphics{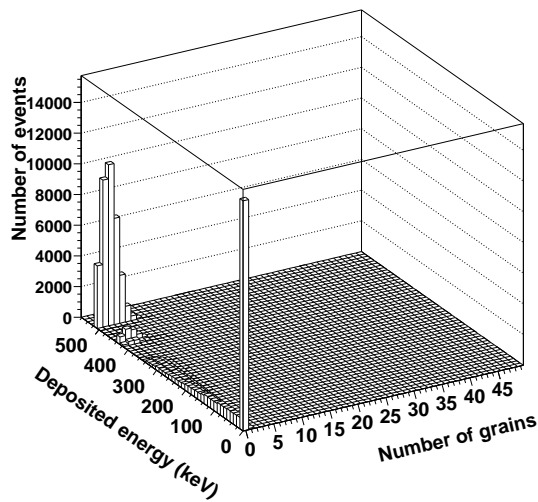}}
 \caption{Number of grains hit  versus the whole energy deposition in the cell (in the litharge and in the grains) \label{figure_5}}
 \end{center}
\end{figure}

This distribution is represented in figure \ref{figure_5} which gives the fraction of concerned grains for a given range of deposited 
energy by an impinging electron. At this stage we 
have a full description of the deposited energies in the suspension.  So far we have 
not used the superconducting state properties. Now we have to estimate the response 
of the superheated superconducting system to this distribution of deposited energy.

\subsection{Response of superheated superconducting microspheres to an energy deposition}

For the purpose of the feasibility study it is enough to limit ourselves to a small 
number of crossed grains thus representing a tiny fraction of their overall number. 
In principle, for each change of state of a grain, because of the long range diamagnetic 
interaction, the whole set of maxima of the equatorial magnetic field is changed. 
However, when only a small number of grains are undergoing a transition one 
can 
assume that the distribution of maxima of the equatorial magnetic field remains the same. 
\begin{figure}[!htb]
 \begin{center}
 \resizebox{!}{7cm}{
 \includegraphics{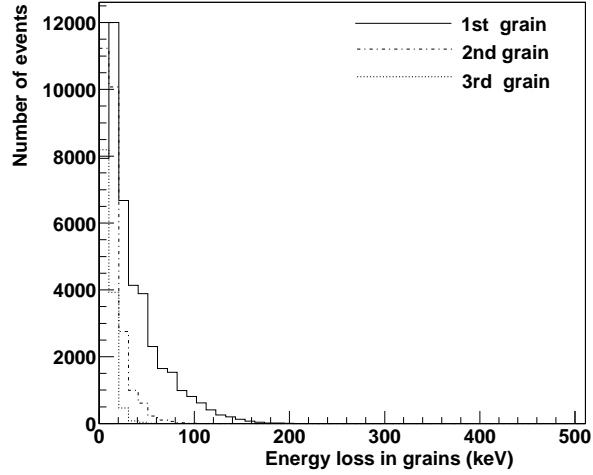}}
 \caption{Energy loss in the grains: the grains are ordered by the amount of deposited energy, the first one is represented by the continuous line \label{figure_6}}
 \end{center}
\end{figure}

As a result the simulation relies on two quantities which, in first approximation 
are taken as independent stochastic phenomena:
\begin{itemize}
\item[-] the energy of the gamma when entering  into the grain. 
\item[-] the equatorial magnetic field on the corresponding grain. 
\end{itemize}
(the impact point is not a relevant parameter because the energy
deposition if large enough always creates an homogeneous nucleation :
there is a uniform and global heating of each flipped grain ( flipped
= crossed grains that will change state)).  In other terms the distribution
of deposited energies among crossed grains is convoluted with the
distribution of their equatorial magnetic field.

\begin{figure}[!htb]
 \begin{center}
 \resizebox{!}{7cm}{
 \includegraphics{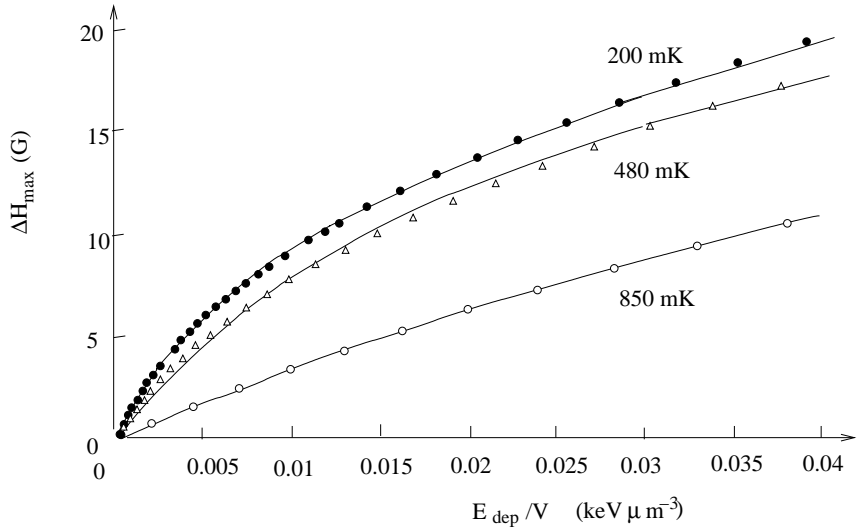}}
 \caption{Energy deposition in the grains versus $\Delta$H$_{max}$. Incidently for each T$_{bath}$
this gives the correspondence between $\Delta$H in gauss and ``$\Delta$H '' in eV. \label{helene}}
 \end{center}
\end{figure}

Figure \ref{helene} (reference \cite{d_helene} page 82) shows the $\Delta$H$_{max}$ in function of the density of energy deposition in the grain, for three different temperatures. If we multiply this density by the volume of the microsphere (the grain) we get for a given temperature, and a certain energy deposition, the range of $\Delta$H, that we can randomly choose between 0 and $\Delta$H$_{max}$.   

\begin{figure}[!htb]
 \begin{center}
 \resizebox{14cm}{14cm}{
 \includegraphics{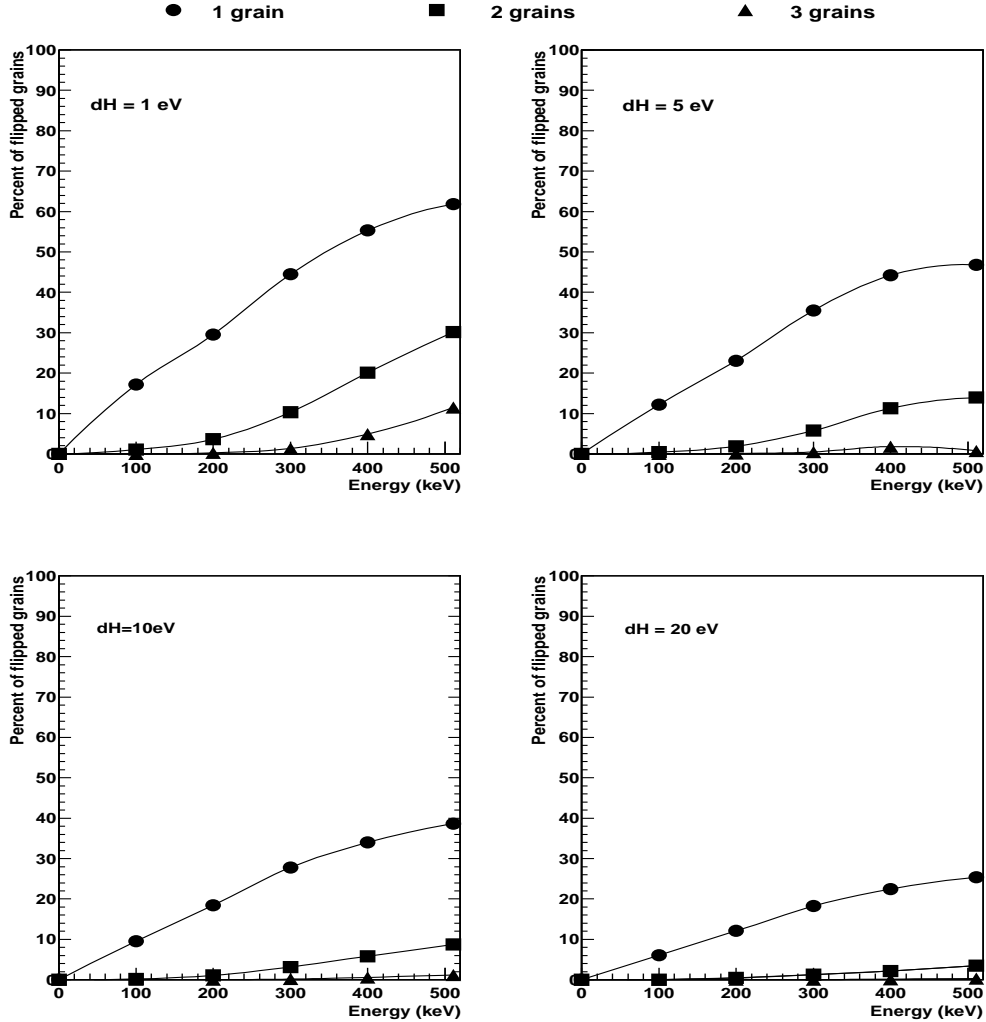}}
 \caption{Photon electronic detection efficiency in function of energy, for grains with 10  micron of  radius  \label{figure_7an}}
 \end{center}
\end{figure}

\begin{figure}[!htb]
 \begin{center}
 \resizebox{14cm}{14cm}{
 \includegraphics{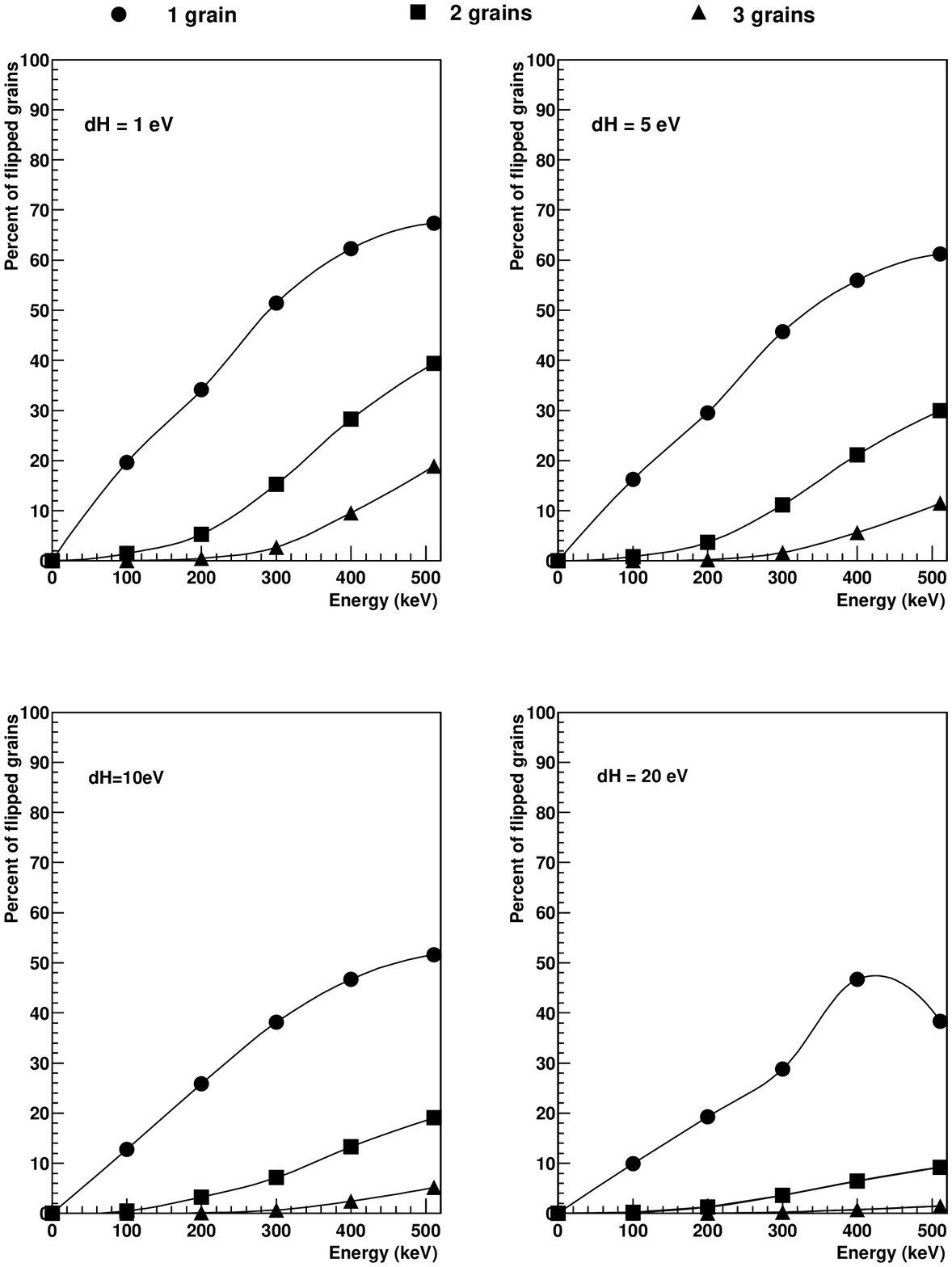}}
 \caption{ Photon electronic detection efficiency in function of energy, for grains with  8 micron of  radius \label{figure_7bn}}
 \end{center}
\end{figure}

\begin{figure}[!htb]
 \begin{center}
 \resizebox{14cm}{14cm}{
 \includegraphics{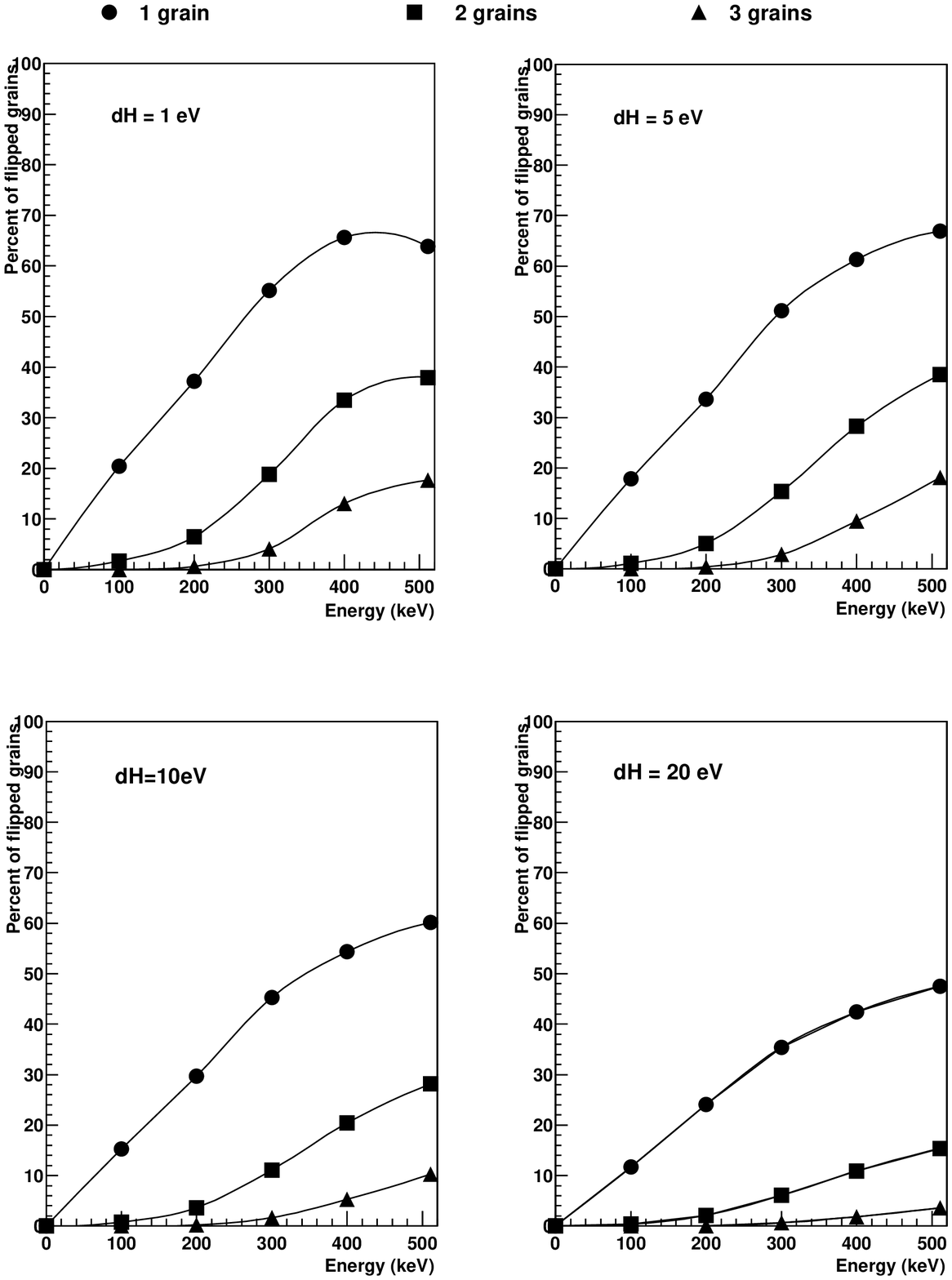}}
 \caption{ Photon electronic detection efficiency in function of energy, for grains with 7 micron of radius \label{figure_7cn}}
 \end{center}
\end{figure}

Thus we obtain as a
function of the impinging photon energy the intrinsic efficiency of
the pixel (figures \ref{figure_7an},\ref{figure_7bn} and \ref{figure_7cn}  )  assuming that at H applied 20\% of the grains are
already flipped to the normal state and that in the interval [H$_{applied}$, 
H$_{applied}$+$\Delta$H$_{max}$] the number of grains per unit of $\Delta$H is
constant.

In fact, on figures \ref{figure_7an}, \ref{figure_7bn}  and  \ref{figure_7cn} 
for a given suspension three efficiencies have been drawn  as a function of the impinging gamma respectively
corresponding to the simultaneous  flipping of 1, 2 or 3 grains. 
3 different radius of grains has been used, 10$\mu$m, 8$\mu$m, 7$\mu$m. Four different $\Delta$H were used, 1 eV, 5 eV, 10 eV, 
and 20 eV per $\mu$m$^{3}$. As
expected the resulting curves are strikingly different  and lead us to
consider that the event selection can be based not only on energy
selection by magnetic field monitoring which is conventional in SSG
systems but, in that case can be combined with signal amplitude
selection. This opens a wide range of configuration for PET. To appreciate the potentialities of a  SSG PET let us
consider the classical case of a small animal camera.
The present state of art of the  electronic read-out allows us to read 1 grain of 10 $\mu$m radius or 2 grains of 8 $\mu$m radius. 5 eV of $\Delta$H is assumed.
$\Delta$H$_{max}$ of 5 eV per cubic micron is a good compromise, allowing to work with reasonable temperature (between 200 and 500 mK, see figure \ref{helene}) and precision (5 keV per $\mu$m$^{3}$ ) correspond to an energy loss of 20 keV in grain of 10 $\mu$m radius, see figure \ref{figure_6}). We can for example estimate the electronic efficiency of the system at 511 keV:

\begin{itemize}

\item

for 10 $\mu$m: 0.534

\item

for 8 $\mu$m: 0.380 

\end{itemize}

A more sophisticated simulation shall mix the different radii of grains, allowing more subtle energy differentiation.

\section{Performance  required for  a small animal PET } 

\begin{figure}[!htb]
 \begin{center}
 \resizebox{!}{7cm}{
 \includegraphics{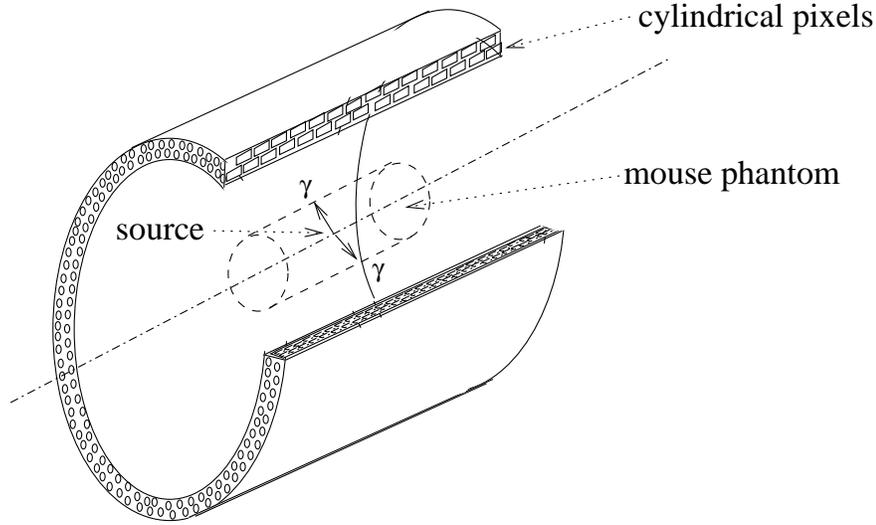}}
 \caption{Micro PET geometry \label{pet}}
 \end{center}
\end{figure}


 The PET requirements were discussed among others by Huber et al
 (\cite{h_huber}). One considers a mouse phantom of 29g placed in the centre of a
 20 cm long cylinder, 5 cm in diameter made of two closed compact layers 
of elementary cylindrical pixels as discussed above (see figure \ref{pet}).

Monte Carlo simulation gives a materialization efficiency (Table 1 ) to 511 Kev
irradiation of the order of 42\% in a pixel. This efficiency can be
higher if one increases the depth of the detecting layer.  This is
possible because, contrary to classical techniques neither the
read-out electronics nor the detecting material are limiting factors
for an efficient design of the whole system.

\begin{figure}[!htb]
 \begin{center}
 \resizebox{!}{7cm}{
 \includegraphics{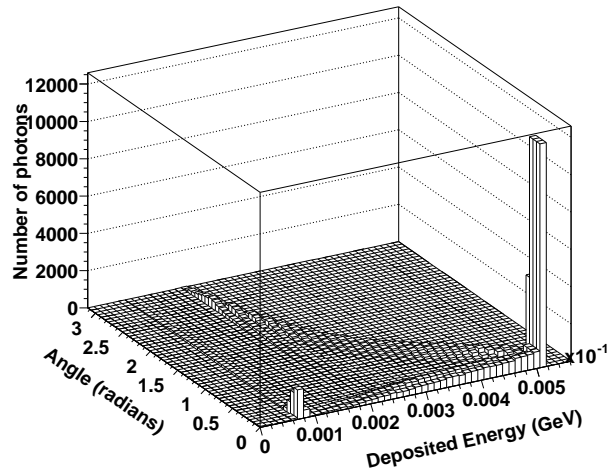}}
 \caption{ Deposited energy in the cell for straight and diffused 
photons, versus the angle between the emitted photon and the hitted cell \label{figure_9}}
 \end{center}
\end{figure}

        The energy deposited in the pixel comes from either a straight gamma (signal) or from a diffused trajectory in the mouse
phantom (noise). Figure  \ref{figure_9} gives the distribution 
of
the deposited energy for the straights  and the diffused
photons. The efficiency given by fig 7 allows an estimation of the
noise equivalent counting rate (NECR) which is the ratio of interest
for PET performances.  For its estimation we follow conventional
notations. The true ($True$) coincidental and scatter ($Scatter$) rates are given by, $\mu$ being a constant,

$$True= \mu \cdot \rho \cdot (\epsilon^{2}) \cdot g_{t} \cdot (P^{2})$$ 

and 

$$Scatter=f_{s} \cdot True $$

respectively, where 

\begin{description}
\item$\rho$ is the activity density, 
\item$\epsilon$ is the single detection efficiency, 
\item$g_{t}$ is a geometry efficiency for true coincidence events, 
\item P is the probability for escape for a 511 keV photon from the small animal, 
\item $f_{s}$ is the scatter fraction. 
\end{description}

The random ($Random$) coincidence event rate is proportional to single event rate ($Single$) 

$$2 \cdot \tau \cdot (Single)^{2}$$ 

where 

$$Single = \rho \cdot \epsilon \cdot g_{r} \cdot P$$

that means:

$$Random = \mu \cdot 2 \cdot \tau \cdot \rho^{2} \cdot \epsilon^{2} \cdot g_{r}^{2} \cdot P^{2} $$

and 

$$Random = True \cdot \rho \cdot g_{t} \cdot 2 \cdot \tau$$ 

where 

\begin{description}
\item
$\tau$ is the coincidence resolving time ( usually 10 or 5 ns) 
\item
$g_{r}$ is the  geometry efficiency for random events, considered the same as $g_{t}^{2}$.
\end{description}

        The Noise Equivalent Count Rate (NECR) is the number of counts
        detected as a function of the activity concentration, after
        correcting the effects of random and scatter events taking
        into account dead-time losses. It is a standard measure of
        signal to noise in reconstructed PET images.

$$NECR= True^{2}/ (True+Scatter+k. Random)$$ 

where 

\begin{description}
\item
k is a correction factor for random coincidences (k=1 or 2).
\end{description}

To illuminate the possibilities of
SSG HDM PET two different grains suspensions and corresponding
electronic tuning are considered: the first one with 10$\mu$m radius
tin grains and with an electronic threshold corresponding to the
magnetic flux penetration in one grain, the second with 8$\mu$m
radius tin grains. We presume working at a temperature (between 200 and 500 mK) allowing a $\Delta$H$_{max}$ of 5 eV. With an electronic threshold corresponding to two
simultaneously flipping grains, the performance for each situation is
given below for $10^{6}$ events, by the following numbers presented in Table 2:

\medskip
\begin{table}
\begin{center}

\begin{tabular}{|l|c|c|c|c|c|}
\hline
 &gamma &gamma & gamma  &readout  &diffused\\ 
 &interaction&direct&diffused&grains &impact\\ 
\hline
20$\mu$m:&420282&319732&100550&170804 &26024\\
\hline 
16$\mu$m:&420282&319732&100550&98587 & 8324\\
\hline
\end{tabular}
\caption{History of 1000000 single photon events}
\end{center}
\end{table}
\medskip

If we suppress the phantom and the depth of the vessel, we get 384990
gamma's directly in our system. Of these 384990, 367770 are on the
511keV as entry energy. In the system with phantom and vessel, we get
267670 photons with 511 keV.  But, in reality, a large fraction of
photons give a signal in the same angle as the direct photons. We consider the threshold angle for diffused gammas as 0.234 radians. 

\begin{description}
\item
The probability of escape P can be evaluated to 0.7281.  
\item
The probability of direct impact is 0.8793.
\item
The solid angle of our system is for 20 cm of length of the detector and 5 cm of radius:

$$4 \pi  (1- \cos ( \arctan(0.5)))$$

$$=4 \pi (1- \frac{2} {\sqrt{5}})$$

and $$g_{r}= \frac {2} {\sqrt{5}}=0.9$$

\item

$\epsilon$ is the product of the materialisation factor by the electronic efficiency which is given by Table 2.

The gamma materialization factor is 

m = 0.367770/0.9=0.419

\begin{table}
\begin{center}
\begin{tabular}{|l|c|c|c|}

\hline


radius          & true          & diffused     & scatter fraction\\ 
\hline
10 $\mu$m & 170804/319732 & 26024/100550 & 26024/170804 \\ 
          & 0.534         & 0.26         &0.1523\\ 
\hline
8 $\mu$m & 98587/319732  & 8324/100550  & 8324/98587\\ 
          & 0.380         & 0.083        & 0.08443\\
\hline
\end{tabular}
\caption{Photon electronic efficiency }
\end{center}

\end{table}

This gives:

For a radius of 10 $\mu$m, $\epsilon$ = 0.419*0.534=0.2036

For a radius of 8 $\mu$m, $\epsilon$ = 0.419*0.380=0.1520

\end{description}

This allows us to estimate the  

$$NECR = True \frac{1}{1+f_{s}+2\rho \cdot g_{t}\cdot \tau}$$

with:

$$\tau=10^{-8}, g_{t}=0.9, P=0.7281$$

and:

$$True=\rho\cdot \epsilon^2 \cdot g_{t} \cdot P^2$$

The maximum acquisition rate is of 10$^8$, this gives, for a source of
 10$^8$ like the one cited in reference \cite{h_huber}:

\begin{description}

\item

For 10 $\mu$m radius,

$$\epsilon = 0.2036,    True=1.98 10^{6}  $$

$$ NECR= \frac{0.0198 10^{8}}{1.1523+\rho 1.8\cdot 10^{-8}}=0.66 \cdot 10^6$$

\item

For 8 $\mu$m radius,

$$\epsilon =  0.1520,  True=9. 10^{5} $$

$$ NECR= \frac{0.009 10 ^{8}}{1.08443+\rho 1.8\cdot 10^{-8}}=0.3 \cdot 10^6$$

\end{description}

The number of diffused, $Scatter$, is 3. 10$^5$ and 0.7 10$^{5}$ respectively in these two cases.

The number of random coincidences, $Random$, is the most important noise:
 with such a source of $10^8$, and a window of 10 nanosecond, we get
 0.9 10$^{6}$ an 0.54 10$^{6}$ respective of noise.  

Improvement of the NECR can be achieved by following paths:

\begin{description}

\item

The $g_{t}$ factor can be increased by a new  geometry covering a larger solid angle.

\item

The NECR is proportional to the square of the materialization factor:
 with three sheets of detecting rows, we should  get an electronic efficiency of 0.784 instead of
 0.4, this implies a multiplication of the NECR of about 4: this should give:

NECR= 2.6 10$^6$ for 10 $\mu$m radius, and NECR= 1.2 10$^6$ for 8 $\mu$m radius. 

\item

Noise reduction is improved with sharper time coincidences: we take 10$^{-8}$ second.

\item

Increasing electronic efficiency can be achieved in
 more sophisticated reconstruction taking into account when two cells give a signal.

 \end{description}

In usual case, sources of lower intensity are used, in this case we get a more efficient NECR (Table 3, for 1 sheet of row):

\begin{table}
\begin{center}
\begin{tabular}{|l|c|c|c|}

\hline


radius          & 10$^{8}$ & 10$^{7}$ & 10$^{6}$   \\ 
\hline
10 $\mu$m & 666000 & 150000 & 17500 \\ 
          & 33\%         & 75\%         &  82\%\\ 
\hline
8 $\mu$m & 300000  & 71000 & 8200\\ 
          & 30\%  &  80\%       & 90\%\\
\hline
\end{tabular}
\end{center}
\caption{NECR according to the intensity of the source (10$^{6}$, 10$^{7}$ and 10$^{8}$ Becquerel), 
and to the radius of the grains, given in counts per second and 
percentage of the whole counting rate.}
\end{table}

\medskip

\section{Conclusions:}
 SSG PET is an alternative to classical techniques pushed
to their limits. As often is the case the flexibility of a new
solution is repaid by the familiarity with new techniques, in this case
very low temperature cryogenics which have made significant progress
toward simplicity and an easy-to-use operation, in the last ten
years.  Low temperature and superconductors are already used for medical imaging (MRI magnets). 
We would underline the fact that SSG PET has no optical component and
no crystal detector. As a result it is open to various designs in spite of the fact
that it must operate at a low temperature. Secondly: individual pixels at
the actual stage of electronics allow high counting rates.  Thirdly :
SSG allows energy selection without any numerical treatment : the
monitoring of the magnetic field is enough.
Although this paper deals with the feasibility of PET systems for small animals we would like also stress that the non crystalline nature of the detector allows the realization of pixels with a volumic fraction at their input in a low density matrix. In that case the detector is mostly sensitive to soft X ray (in fact it was most often used in that range of energy). Therefore a combined PET+CT system with the same geometry and electronics is perfectly feasible.

\section{Acknowledgments}
We would like to thank MM Bonnierbale for his technical help. We would like to thank M Lallemand, Director of ASCI (Application Scientifique du Calcul Intensif), for his support and help with the numerical simulations.

\bibliography{granus}
\end{document}